\documentclass[letterpaper, 10pt, conference]{ieeeconf}  

\IEEEoverridecommandlockouts                              

\usepackage[pdftex]{graphicx}
\usepackage{url,cite}

\title{\LARGE \bf
Student Teaching and Research Laboratory Focusing on Brain--computer Interface Paradigms \\ -- A Creative Environment for Computer Science Students --
}

\author{Tomasz M. Rutkowski$^{1}$
\thanks{$^*$The presented student research and educational laboratory was supported in part by the Strategic Information and Communications R\&D Promotion Program (SCOPE) no. 121803027 of The Ministry of Internal Affairs and Communication in Japan, YAMAHA Corporation and by KAKENHI, the Japan Society for the Promotion of Science, grant no.~$24243062.$}
\thanks{$^{1}$Tomasz M. Rutkowski with the Department of Computer Science and Life Science Center of TARA, University of Tsukuba, 1-1-1 Tennodai, Tsukuba, Ibaraki, 305-8577 Japan,
        {\tt\small tomek@bci-lab.info} {\tt\small http://bci-lab.info/}}}%

\begin{document}

\maketitle
\thispagestyle{empty}
\pagestyle{empty}

\begin{abstract}
This paper presents an applied concept of a brain--computer interface (BCI) student research laboratory (BCI--LAB) at the Life Science Center of TARA, University of Tsukuba, Japan. Several successful case studies of the student projects are reviewed together with the BCI Research Award 2014 winner case. The BCI--LAB design and project--based teaching philosophy is also explained. Future teaching and research directions summarize the review.
\end{abstract}

\section{INTRODUCTION}

Brain computer interface (BCI) is a technology that uses brain neural activities only to control a computer, or a machine, without any body muscle movements~\cite{bciBOOKwolpaw}. This technology could allow disabled people, such as locked--in syndrome (LIS) suffering patients~\cite{lis139review1986}, to regain communication skills or to manage daily life support related functions. 
On the other hand, the BCI technology creates a perfect platform for computer science students to study computational neuroscience--based neurotechnology methods. There are also many types of BCIs based on the utilized brainwave signals and experimental paradigms. This allows to create the multi--sensory experimental labs with biomedical (e.g. neurophysiological) signals capturing and processing in mind. Creation of multi--sensory stimuli allows also the students to practically develop and test multimedia environments for a future deployment of the developed paradigms for paralyzed patients. This review paper presents a successful implementation of the BCI teaching and experimental lab at the Life Science Center of TARA, University of Tsukuba, Japan. The experimental focus of the non--invasive BCI lab is on the external stimulus--driven paradigms relaying on event related potential's (ERP) P300 responses~\cite{bciBOOKwolpaw} allowing for multi--sensory stimulation application. The sensory modalities available for experimenting are as follows,
(i) spatial auditory: with multi--loudspeaker studio configurations~\cite{nozomuANDtomekAPSIPA2013,MoonJeongBCImeeting2013,tomekNER2013,bertrandSCIS2014,moonjeongPROCEDIA2014,tomekICCN2015} and headphone viral sound--based~\cite{chisakiAEARU2015,hrirBCIconf2014}; 
(ii) spatial visual: P300~\cite{MoonJeongBCImeeting2013} and steady--state response--based~\cite{daikiAEARU2015,aminakaAPSIPA2014}; 
(iii) spatial tactile: realized with vibration--based P300~\cite{tbBCIconf2014,yajimaAPSIPA2014,tomekJNM2015}, tactile--force~\cite{tomekMTA2014}, pin--pressure~\cite{kensukeSCIS2014}, and airborne ultrasonic tactile display (AUTD) contactless somatosensory modality~\cite{tomekCBMI2015,autdBCIconf2014}.
Non--invasive BCIs are the most safe because the brainwave sensors are usually attached to the surface of a human head, thus such experiments, within the institutional ethical committee guidelines, are perfectly suitable for computer science students. All the experiments reviewed and reported in this paper were conducted 
in agreement with the ethical committee guidelines of the Faculty of Engineering, Information and Systems at University of Tsukuba, Tsukuba, Japan (experimental permission no.~$2013R7$). 
The non--invasive BCI modalities, however, suffer usually from lower brainwave quality due to a very poor signal to noise ratio comparing to the invasive systems, which creates a perfect educational project platform for the engineering students, who design filtering and feature extraction algorithms to cope with the above problems~\cite{yoshihiroANDtomekAPSIPA2012,yoshihiroANDtomekAPSIPA2013,tomekJNM2015}. The non--invasive BCIs are divided into stimulus--driven.
The stimulus--driven BCIs constitute the interactive interfacing solutions in which the commands to operate a computer or an application are associated with sensory stimuli delivered to the user. The user generates a signature brainwave captured in EEG by focusing attention (on the so--called \emph{target} eliciting the P300) or by ignoring (the \emph{non--target}) the randomly presented sensory auditory, tactile or visual stimuli. The students of the BCI--LAB design those spatial sensory stimulations and enhance practically their multimedia knowledge with support of the neurotechnology (brain response--driven) applications.  

Among the recently developed BCIs the visual modality~\cite{bciBOOKwolpaw} is currently the most successful because the evoked P300 responses have usually the largest amplitudes allowing for the most easy classification~\cite{MoonJeongBCImeeting2013}. However, it is known that LIS users (e.g. advanced stage amyotrophic lateral sclerosis patients) cannot use visual modality because they gradually lose intentional muscle control, including eye movements, focusing or intentional blinking. Our research hypothesis is that the auditory and tactile modalities BCI shall be a more suitable solution for communication establishment for LIS users. The students of the BCI--LAB have an opportunity to design   ``novel sensory stimulations'' by combining tactile and bone--conduction--auditory~\cite{HiromuBCImeeting2013}; audiovisual~\cite{MoonJeongBCImeeting2013}; and contactless tactile~\cite{tomekCBMI2015,autdBCIconf2014} paradigms.

The rest of the paper is organized as follows. In the next section we describe the teaching philosophy behind the BCI--LAB concept. Next the experimental environment is described focusing on the hardware, software and stimulus generators. Finally the already published student projects are briefly described. Finally, the conclusions are drawn and future research and educational directions outlined.

\section{TEACHING PHILOSOPHY}

A teaching philosophy implemented in the BCI--LAB is based on illustrating the beauty of a neuroscienceÑ-driven creative innovation of media and human augmentation. Even though computational neuroscience and the modern media or human augmentation might be regarded as not very closely related research disciplines in the eyes of the general public, practical approaches often result with the both disciplines sharing outcomes from the both fields. For most the BCI--LAB students the projects are their first chance to study ÒbrainÑ-relatedÓ subjects in application to realÑ-life multimedia and human augmentation applications' development. To begin each new project, we use in the lab an analogy to real life examples to lead the students to the reasonability of the method. It happens that when we talk about the vision, hearing or some other perceptual methods we are ourselves so familiar with, we find ourselves facing a room of confused faces. The best remedy for such initial confusion is a participation in as experimental subjects in ongoing lab projects usually run by the senior students. A master--student relationship between students could be simply realized, which helps with lab culture establishment. 
A base for an successful experimental lab foundation is an interactive and student--friendly teaching approach. No lab seminar notes can be recycled from one semester to the next, nor is it best to use the same examples. Students are different and their ways of developing understanding too. Our lab philosophy is that a senior experimenter should always be ready to adjust an advisory strategy due to the needs of the new students. If any new to the project student complains about the problem or a lack of available solution, a drill experimental example may become necessary; if there is one tricky problem in the experimental setting, a little hint given will make the studentsÕ efforts more inspired; if the experimental design has a dull nature, adding some interesting trial examples may lighten it up and make the lab life more enjoyable and less slumberous. 

The concept of our BCI--LAB is to train the computer science students in computational neuroscience problems by using "hands--on" practical problems' learning philosophy.

\section{BCI--LAB ENVIRONMENT}

There are three pillars of a successful experimental stimulus--driven BCI--LAB, namely a good experimental hardware environment; flexible software platforms for brainwave signal processing and classification; an excellent multi--sensory stimulus design platform. 

The experimental hardware, allowing for EEG capture for the online BCI experiments discussed in the following sections, has been set up based on g.USBamp and g.MOBIlab+ from g.tec Medical Engineering, Austria (used in the majority of the projects in BCI--LAB), as well as on vAmp from Brain Products GmbH, Germany (used only in ~\cite{moonjeongPROCEDIA2014,tomekICCN2015}). All the systems used the active EEG electrodes g.LADYbird and g.SAHARA from g.tec Medical Engineering, Austria. Due to different EEG amplifiers manufacturers we decided to use, and to extend with own functions, flexible signal acquisition and classification environments BCI2000~\cite{bci2000book} and OpenVibe~\cite{openvibe}. The both environments allow for flexible extensions also developed and applied by our team~\cite{yoshihiroANDtomekAPSIPA2013,daikiAEARU2015}.
Finally the spatial auditory, visual and tactile stimulation generators were programmed by our team in MAX~6~\cite{maxMSP} and delivered to the users via multichannel sound cards, large computer displays or tactile transducers managed by ARDUINO micro--controller boards.

\section{BCI--LAB STUDENT PROJECTS}

\subsubsection{Multi--loudspeaker--based Spatial Auditory BCIs}

\begin{figure}[!t]
	\vspace{0.2cm}
	\begin{center}
  	\includegraphics[width=0.7\linewidth]{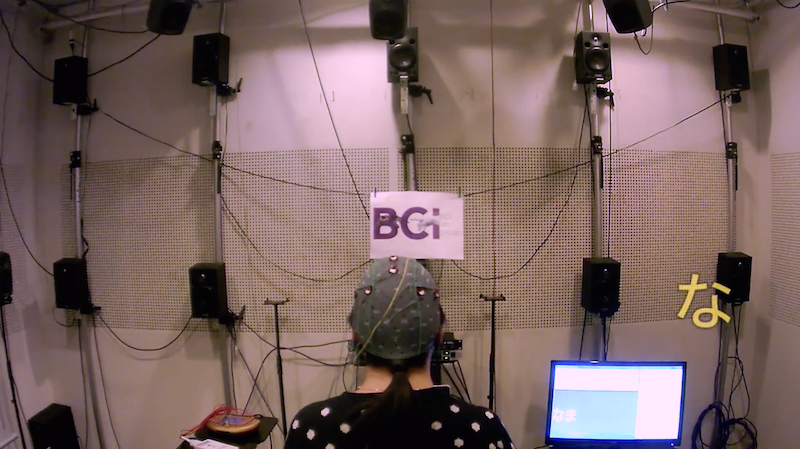}
 	\end{center}
	\vspace{-0.2cm}
 	\caption{A photograph of a two--step input Japanese kana characters spatial auditory BCI~\cite{moonjeongPROCEDIA2014,tomekICCN2015} project realized in the BCI--LAB. The user in the photograph sits in front of ten loudspeakers (sponsored by YAMAHA Corporation) delivering the spatial sound images. The yellow Japanese ``na'' character visualizes (not seen to the user) the sound image location. An online video from the experiment is available at~\cite{youtubeBCIspellerMOONJEONG}.}\label{fig:moonjeongBCI}
	\vspace{-0.4cm}
\end{figure}

Three different multi--loudspeaker--based spatial auditory BCI paradigms were realized in the BCI--LAB allowing the students to design and test multi--source auditory environments using MAX~6 visual programming environment~\cite{maxMSP}. The first approach implemented a vector--based amplitude panning method to spatial horizontal sound images positioned octagonally around the user's head~\cite{nozomuANDtomekAPSIPA2013}. The second approach involved moving sound stimuli realized in half--spherical loudspeaker environments~\cite{tomekNER2013,bertrandSCIS2014}. The third, and the most successful application, realized a two--step--input speller based on the spatial auditory sound images presentation~\cite{moonjeongPROCEDIA2014,tomekICCN2015} as shown in Figure~\ref{fig:moonjeongBCI}. The application has been developed for $45$  Japanese syllabary (\emph{kana}) letters using a two--step--input procedure in order to create an easy to use BCI--speller interface. The results of the conducted online EEG experiments have shown the feasibility of the Japanese \emph{kana} characters set in two--step--input saBCI--speller paradigm. Also we confirmed that the accuracy of each modality had no significant differences. A video documenting  the online saBCI--speller experiment is available at~\cite{youtubeBCIspellerMOONJEONG}. The three presented  auditory BCI paradigms allowed for training of the computer science students in spatial sound design and biomedical signal acquisition (EEG), processing and classification for the online applications.

\subsubsection{Headphone--based Spatial Auditory BCI Paradigm}

This study reported on a head related impulse response (HRIR) application to an auditory spatial brain-computer interface (BCI) speller paradigm~\cite{hrirBCIconf2014,chisakiAEARU2015}. Six experienced and five BCI--naive users participated in an experimental spelling set up based on five Japanese vowels. Obtained auditory evoked potentials resulted with encouragingly good and stable P300--responses in online BCI experiments. Our case study indicated that the headphone reproduced auditory (HRIR--based) spatial sound paradigm could be a viable alternative to the established multi--loudspeaker surround sound BCI--speller applications, as far as healthy pilot study users are concerned. This project allowed the students to practically apply headphone--based spatial sound virtualization to the online auditory BCI application~\cite{hrirBCIconf2014,chisakiAEARU2015}.

\subsubsection{Visual SSVEP and cVEP Paradigms}

\begin{figure}[!t]
	\vspace{0.2cm}
	\begin{center}
  	\includegraphics[width=0.6\linewidth]{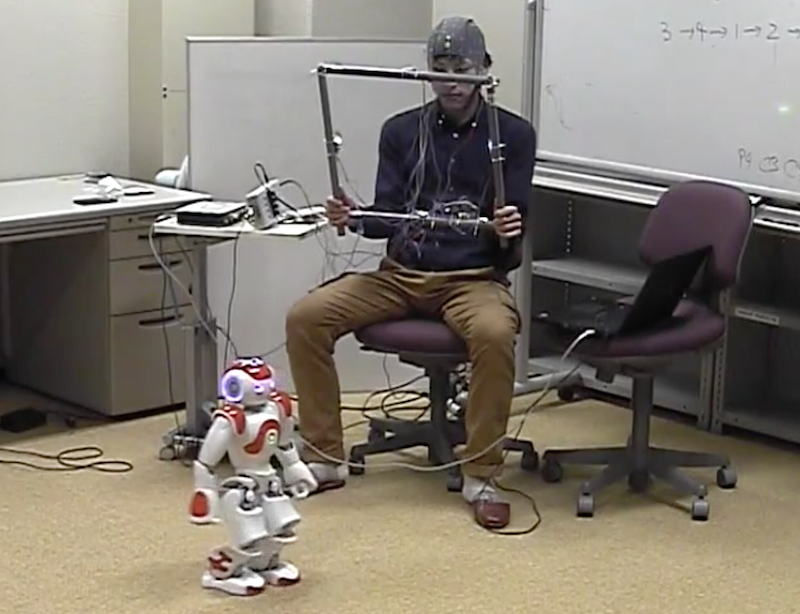}
 	\end{center}
	\vspace{-0.2cm}
 	\caption{A photograph of a chromatic SSVEP BCI project~\cite{aminakaAPSIPA2014,daikiAEARU2015} realized in the BCI--LAB. The user in the photograph sits in front of a frame with four green--blue LEDs. A humanoid robot is managed using four commands from the SSVEP BCI. An online video from the experiment is available at~\cite{youtubeSSVEPBCInao}.}\label{fig:daikiBCI}
	\vspace{-0.5cm}
\end{figure}

A novel approach to steady--state visual evoked potentials (SSVEP) based BCI was developed~\cite{aminakaAPSIPA2014,daikiAEARU2015} with further extension to a code--modulated visual evoked potential (cVEP) approach. To minimize possible side effects of the monochromatic light SSVEP we proposed to utilize color information in form of the green and blue interlaced flicker (see Figure~\ref{fig:daikiBCI} with a robot control application). The feasibility of the proposed method was evaluated in a comparison of the classical monochromatic versus the proposed colorful SSVEP responses. The proposed method scored with higher accuracies comparing to the monochromatic SSVEP~\cite{daikiAEARU2015}. This visual modality project allowed the student to attack the classical problems related to flickering light--based stimuli (danger of photosensitive epilepsy, etc.). As result two solutions were proposed based on higher frequency and cVEP--based stimuli together with less sensitive chromatic light modality.

\subsubsection{Bone--conduction Auditory and Tactile BCI Paradigm}

\begin{figure}[!t]
	\vspace{0.2cm}
	\begin{center}
  	\includegraphics[width=0.7\linewidth]{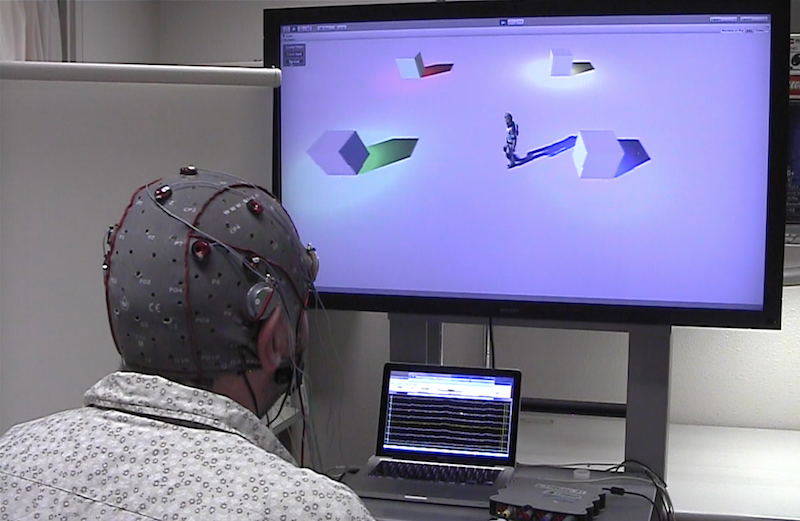}
 	\end{center}
	\vspace{-0.2cm}
 	\caption{A photograph of a bone--conduction auditory and tactile multi--sensory BCI Paradigm~\cite{HiromuBCImeeting2013,tomekJNM2015} project realized in the BCI--LAB. The user in the photograph manages a walking virtual avatar based on intentional responses to the combined tactile and bone--conduction auditory stimuli delivered thorough vibrotactile transducers embedded in the EEG cap (large metallic discs above the ear and on a forehead in the picture).}\label{fig:hiromuBCI}
	\vspace{-0.4cm}
\end{figure}

In this project we studied the extent to which vibrotactile stimuli delivered to the head of a subject could serve as a platform for the BCI paradigm~\cite{HiromuBCImeeting2013} (see Figure~\ref{fig:hiromuBCI}). Six head positions were used to evoke combined somatosensory and auditory (via bone-conduction effect) brain responses, in order to define a multimodal tactile and auditory brain computer interface (taBCI). Experimental results on subjects performing online taBCI, using stimuli with a moderately fast inter-stimulus-interval, validated the taBCI paradigm, while the feasibility of the concept was illuminated through encourafing information-transfer rate case studies. In the reported experiments~\cite{HiromuBCImeeting2013} only a single BCI-na•ve subject obtained $100\%$ and also one obtained $0\%$ for the six digit sequence spelling accuracy with 5-trials averaging procedure. The results were further improved with an implementation of a novel synchro--squeezing EEG preprocessing method as reported in~\cite{tomekJNM2015}. This project allowed the BCI--LAB students for a creation of a novel multi--sensory BCI platform~\cite{HiromuBCImeeting2013} together with the new signal processing method development~\cite{tomekJNM2015}.

\subsubsection{Vibrotactile and Tactile--pressure BCI Paradigms}

A series of successful vibrotactile~\cite{tbBCIconf2014,yajimaAPSIPA2014} and tactile--pressure/force~\cite{kensukeSCIS2014,tomekMTA2014} paradigms were developed with the BCI--LAB approaching a problem of communication alternatives design for those LIS patients with no vision or functional hearing (due to the so called ear stacking syndrome). In these spatial tactile projects the contributing students combined successfully their knowledge of multichannel stimulus generation systems development in MAX~6 environment~\cite{maxMSP} together with brainwave processing and classification.

\subsubsection{Airborne Ultrasonic Tactile Display BCI Paradigm}

\begin{figure}[!t]
	\vspace{0.2cm}
	\begin{center}
  	\includegraphics[width=0.6\linewidth]{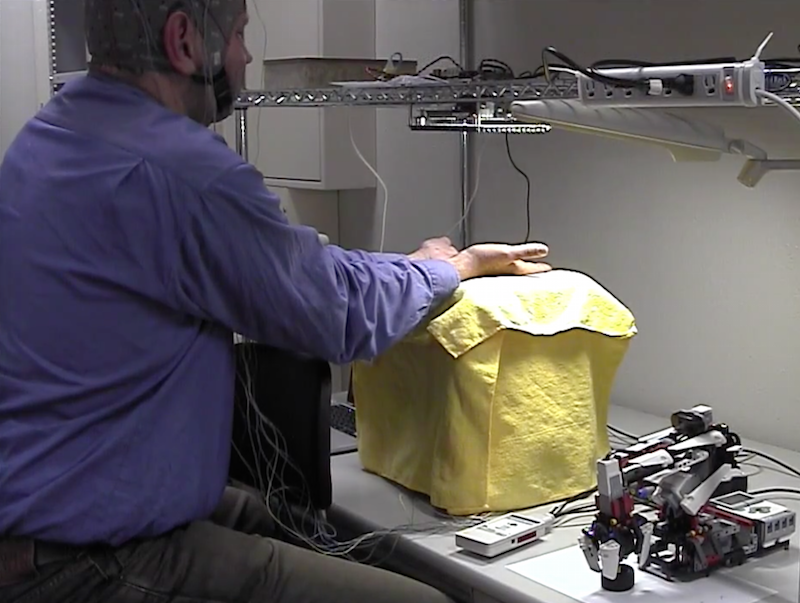}
 	\end{center}
	\vspace{-0.2cm}
 	\caption{A photograph of the AUTD BCI Paradigm~\cite{autdBCIconf2014,tomekCBMI2015} project realized in the BCI--LAB. The user in the photograph manages a small LEGO robot based on intentional responses to the contactless tactile stimuli delivered thorough ultrasonic beam generated by the AUTD. An online video from the experiment is available at~\cite{youtubeAUTDbciROBOT}.}\label{fig:autdBCI}
	\vspace{-0.4cm}
\end{figure}

In this BCI Research Award 2014 winning project a contact--less somatosensory stimuli were delivered via an airborne ultrasonic tactile display (AUTD) to the palms of a user (see Figure~\ref{fig:autdBCI}) and served as a platform for the BCI paradigm (autdBCI)~\cite{autdBCIconf2014,tomekCBMI2015}. Six palm positions were used to evoke combined somatosensory brain responses, in order to define a novel contact-less tactile BCI. A comparison was made with classical attached vibrotactile transducers (tBCI). Experiment results of $13$ subjects performing online experiments validated the novel BCI paradigm. 
%
In the case of the autdBCI, only a single user's results were bordering on the level of chance, and four subjects attained $100\%$ ($10$ trials averaging). This project was a result of a collaboration of two laboratories~\cite{autdBCIconf2014,tomekCBMI2015} allowing to train students in a novel contactless tactile modality and BCI online application development.

\section{CONCLUSIONS} 

The reviewed successful student projects in this paper confirmed a hypothesis of the computational neuroscience (neurotechnology) topics teaching suitability for the computer science students based on the project--based education principles realized in the BCI--LAB by introducing the modern signal processing, machine learning, multimedia (multi--sensory stimulation) and neuroscience topics in a single approach to the new BCI experimental paradigms developments. 

The future educational concepts development shall include interaction with real patients in need since so far the presented BCI--LAB projects were tested with healthy users.


\begin{thebibliography}{10}
\providecommand{\url}[1]{#1}
\csname url@samestyle\endcsname
\providecommand{\newblock}{\relax}
\providecommand{\bibinfo}[2]{#2}
\providecommand{\BIBentrySTDinterwordspacing}{\spaceskip=0pt\relax}
\providecommand{\BIBentryALTinterwordstretchfactor}{4}
\providecommand{\BIBentryALTinterwordspacing}{\spaceskip=\fontdimen2\font plus
\BIBentryALTinterwordstretchfactor\fontdimen3\font minus
  \fontdimen4\font\relax}
\providecommand{\BIBforeignlanguage}[2]{{%
\expandafter\ifx\csname l@#1\endcsname\relax
\typeout{** WARNING: IEEEtran.bst: No hyphenation pattern has been}%
\typeout{** loaded for the language `#1'. Using the pattern for}%
\typeout{** the default language instead.}%
\else
\language=\csname l@#1\endcsname
\fi
#2}}
\providecommand{\BIBdecl}{\relax}
\BIBdecl

\bibitem{bciBOOKwolpaw}
J.~Wolpaw and E.~W. Wolpaw, Eds., \emph{Brain-Computer Interfaces: Principles
  and Practice}.\hskip 1em plus 0.5em minus 0.4em\relax Oxford University
  Press, 2012.

\bibitem{lis139review1986}
J.~R. Patterson and M.~Grabois, ``Locked-in syndrome: a review of 139 cases.''
  \emph{Stroke}, vol.~17, no.~4, pp. 758--64, 1986. 
  
\bibitem{nozomuANDtomekAPSIPA2013}
N.~Nishikawa, S.~Makino, and T.~M. Rutkowski, ``Spatial auditory bci paradigm
  based on real and virtual sound image generation,'' in \emph{Signal and
  Information Processing Association Annual Summit and Conference (APSIPA),
  2013 Asia-Pacific}, 2013, pp. 1--5, paper ID 387. 
  
\bibitem{MoonJeongBCImeeting2013}
M.~Chang, N.~Nishikawa, Z.~R. Struzik, K.~Mori, S.~Makino, D.~Mandic, and T.~M.
  Rutkowski, ``Comparison of {P300} responses in auditory, visual and
  audiovisual spatial speller {BCI} paradigms,'' in \emph{Proceedings of the
  Fifth International Brain-Computer Interface Meeting 2013}.\hskip 1em plus
  0.5em minus 0.4em\relax Asilomar Conference Center, Pacific Grove, CA USA:
  Graz University of Technology Publishing House, Austria, June 3-7, 2013, p.
  Article ID: 156. 

\bibitem{tomekNER2013}
Y.~Lelievre and T.~M. Rutkowski, ``Novel virtual moving sound-based spatial
  auditory brain-computer interface paradigm,'' in \emph{Neural Engineering
  (NER), 2013 6th International IEEE/EMBS Conference on}.\hskip 1em plus 0.5em
  minus 0.4em\relax IEEE Engineering in Medicine and Biology Society, 2013, pp.
  9--12, arXiv:1308.2630 http://arxiv.org/abs/1308.2630. 
  
\bibitem{bertrandSCIS2014}
B.~Hieronymus, H.~Mori, and T.~M. Rutkowski, ``Brain-computer interface using
  ambisonics-reproduced dynamic sound image effects,'' in \emph{Soft Computing
  and Intelligent Systems (SCIS), 2014 Joint 7th International Conference on
  and Advanced Intelligent Systems (ISIS), 15th International Symposium on},
  December 2014, pp. 301--306. 

\bibitem{moonjeongPROCEDIA2014}
M.~Chang, K.~Mori, S.~Makino, and T.~M. Rutkowski, ``Spatial auditory two-step
  input Japanese syllabary brain-computer interface speller,'' \emph{Procedia
  Technology}, vol.~18, pp. 25 -- 31, 2014. 
\bibitem{tomekICCN2015}
M.~Chang and T.~M. Rutkowski, ``Two--step input japanese kana characters
  spatial auditory brain--computer interface,'' in \emph{Advances in Cognitive
  Neurodynamics Volume (V)}.\hskip 1em plus 0.5em minus 0.4em\relax Springer
  Netherlands, 2015, pp. (accepted, in press).

\bibitem{chisakiAEARU2015}
C.~Nakaizumi, T.~Matsui, K.~Mori, S.~Makino, and T.~M. Rutkowski, ``Spatial
  auditory brain-computer interface using head related impulse response,'' in
  \emph{Proceedings of The 10th {AEARU} Workshop on Computer Science and Web
  Technologies (CSWT-2015)}.\hskip 1em plus 0.5em minus 0.4em\relax University
  of Tsukuba, February 2015, pp. 37--38.

\bibitem{hrirBCIconf2014}
------, ``Head--related impulse response-based spatial auditory brain--computer
  interface,'' in \emph{Proceedings of the 6th International Brain-Computer
  Interface Conference 2014}, G.~Mueller-Putz, G.~Bauernfeind, C.~Brunner,
  D.~Steyrl, S.~Wriessnegger, and R.~Scherer, Eds.\hskip 1em plus 0.5em minus
  0.4em\relax Graz University of Technology Publishing House, 2014, pp. Article
  ID 020--1--4.

\bibitem{daikiAEARU2015}
D.~Aminaka, S.~Makino, and T.~M. Rutkowski, ``{SSVEP} brain-computer interface
  using green and blue lights,'' in \emph{Proceedings of The 10th {AEARU}
  Workshop on Computer Science and Web Technologies (CSWT-2015)}.\hskip 1em
  plus 0.5em minus 0.4em\relax University of Tsukuba, February 2015, pp.
  39--40.

\bibitem{aminakaAPSIPA2014}
------, ``Chromatic {SSVEP BCI} paradigm targeting the higher frequency eeg
  responses,'' in \emph{Asia-Pacific Signal and Information Processing
  Association, 2014 Annual Summit and Conference (APSIPA)}, December 2014, pp.
  1--7. 

\bibitem{tbBCIconf2014}
T.~Kodama, S.~Makino, and T.~M. Rutkowski, ``Spatial tactile brain-computer
  interface paradigm applying vibration stimuli to large areas of user's
  back,'' in \emph{Proceedings of the 6th International Brain-Computer
  Interface Conference 2014}, G.~Mueller-Putz, G.~Bauernfeind, C.~Brunner,
  D.~Steyrl, S.~Wriessnegger, and R.~Scherer, Eds.\hskip 1em plus 0.5em minus
  0.4em\relax Graz University of Technology Publishing House, 2014, pp. Article
  ID 032--1--4. 

\bibitem{yajimaAPSIPA2014}
H.~Yajima, S.~Makino, and T.~M. Rutkowski, ``P300 responses classification
  improvement in tactile bci with touch-sense glove,'' in \emph{Asia-Pacific
  Signal and Information Processing Association, 2014 Annual Summit and
  Conference (APSIPA)}.\hskip 1em plus 0.5em minus 0.4em\relax IEEE, December
  2014, pp. 1--7. 

\bibitem{tomekJNM2015}
T.~M. Rutkowski and H.~Mori, ``Tactile and bone-conduction auditory brain
  computer interface for vision and hearing impaired users,'' \emph{Journal of
  Neuroscience Methods}, vol. 244, no.~0, pp. 45 -- 51, 2015, brain Computer
  Interfaces; Tribute to Greg A. Gerhardt. 
  
\bibitem{tomekMTA2014}
S.~Kono and T.~M. Rutkowski, ``\BIBforeignlanguage{English}{Tactile-force
  brain-computer interface paradigm},''
  \emph{\BIBforeignlanguage{English}{Multimedia Tools and Applications}}, vol.
  'Online First' on SpringerLink, pp. 1--13, December 2014. 
  
\bibitem{kensukeSCIS2014}
K.~Shimizu, H.~Mori, S.~Makino, and T.~M. Rutkowski, ``Tactile pressure
  brain-computer interface using point matrix pattern paradigm,'' in \emph{Soft
  Computing and Intelligent Systems (SCIS), 2014 Joint 7th International
  Conference on and Advanced Intelligent Systems (ISIS), 15th International
  Symposium on}, December 2014, pp. 473--477. 
  
\bibitem{tomekCBMI2015}
T.~M. Rutkowski and H.~Shinoda, ``Airborne ultrasonic tactile display
  contactless brain-computer interface paradigm,'' in \emph{2015 International
  Workshop on Clinical Brain-Machine Interface Systems (CBMI 2015)}, Tokyo,
  Japan, March 13--15, 2015, p.~39.

\bibitem{autdBCIconf2014}
K.~Hamada, H.~Mori, H.~Shinoda, and T.~M. Rutkowski, ``Airborne ultrasonic
  tactile display brain-computer interface paradigm,'' in \emph{Proceedings of
  the 6th International Brain-Computer Interface Conference 2014},
  G.~Mueller-Putz, G.~Bauernfeind, C.~Brunner, D.~Steyrl, S.~Wriessnegger, and
  R.~Scherer, Eds.\hskip 1em plus 0.5em minus 0.4em\relax Graz University of
  Technology Publishing House, 2014, pp. Article ID 018--1--4. 
  
\bibitem{yoshihiroANDtomekAPSIPA2012}
Y.~Matsumoto, N.~Nishikawa, S.~Makino, T.~Yamada, and T.~Rutkowski, ``Auditory
  steady-state response stimuli based {BCI} application - the optimization of
  the stimuli types and lengths,'' in \emph{Signal Information Processing
  Association Annual Summit and Conference (APSIPA ASC), 2012 Asia-Pacific},
  2012, pp. 1--7, paper ID 285. 

\bibitem{yoshihiroANDtomekAPSIPA2013}
Y.~Matsumoto, S.~Makino, K.~Mori, and T.~M. Rutkowski, ``Classifying {P300}
  responses to vowel stimuli for auditory brain-computer interface,'' in
  \emph{Signal and Information Processing Association Annual Summit and
  Conference (APSIPA), 2013 Asia-Pacific}, 2013, pp. 1--5, paper ID 388.

\bibitem{HiromuBCImeeting2013}
H.~Mori, Y.~Matsumoto, Z.~R. Struzik, K.~Mori, S.~Makino, D.~Mandic, and T.~M.
  Rutkowski, ``Multi-command tactile and auditory brain computer interface
  based on head position stimulation,'' in \emph{Proceedings of the Fifth
  International Brain-Computer Interface Meeting 2013}.\hskip 1em plus 0.5em
  minus 0.4em\relax Asilomar Conference Center, Pacific Grove, CA USA: Graz
  University of Technology Publishing House, Austria, June 3-7, 2013, p.
  Article ID: 095.

\bibitem{bci2000book}
G.~Schalk and J.~Mellinger, \emph{A Practical Guide to Brain--Computer
  Interfacing with {BCI2000}}.\hskip 1em plus 0.5em minus 0.4em\relax
  Springer-Verlag London Limited, 2010.

\bibitem{openvibe}
Y.~Renard, F.~Lotte, G.~Gibert, M.~Congedo, E.~Maby, V.~Delannoy, O.~Bertrand,
  and A.~L{\'e}cuyer, ``Openvibe: an open-source software platform to design,
  test, and use brain-computer interfaces in real and virtual environments,''
  \emph{Presence: teleoperators and virtual environments}, vol.~19, no.~1, pp.
  35--53, 2010.

\bibitem{maxMSP}
\BIBentryALTinterwordspacing
``Max 6,'' 2012. [Online]. Available: \url{http://cycling74.com/}
\BIBentrySTDinterwordspacing

\bibitem{youtubeBCIspellerMOONJEONG}
\BIBentryALTinterwordspacing
T.~M. Rutkowski, ``Japanese kana character spatial auditory {BCI} speller.''
  [Online]. Available: \url{http://youtu.be/BY1kSqh1NEY}
\BIBentrySTDinterwordspacing

\bibitem{youtubeSSVEPBCInao}
\BIBentryALTinterwordspacing
------, ``Chromatic {SSVEP BCI-based NAO} robot control.'' [Online]. Available:
  \url{http://youtu.be/gkzkJ7VM5ps}
\BIBentrySTDinterwordspacing

\bibitem{youtubeAUTDbciROBOT}
\BIBentryALTinterwordspacing
------, ``The {autdBCI} and a robot control (the winner project of {The BCI
  Annual Research Award 2014}),'' {YouTube} video. [Online]. Available:
  \url{http://youtu.be/JE29CMluBh0}
\BIBentrySTDinterwordspacing

\end{thebibliography}


\addtolength{\textheight}{-12cm}   


\end{document}